\documentclass[10pt,conference]{IEEEtran}
\usepackage{todonotes}
\usepackage{amssymb}

\ifCLASSINFOpdf
\else
\fi
\usepackage{algorithm}
\usepackage[noend]{algpseudocode}
\usepackage{algorithmicx}
\ifCLASSOPTIONcompsoc
  \usepackage[caption=false,font=normalsize,labelfont=sf,textfont=sf]{subfig}
\else
  \usepackage[caption=false,font=footnotesize]{subfig}
\fi
\usepackage{url}

\begin{document}
%
\title{Stacked Thompson Bandits}

\author{\IEEEauthorblockN{Lenz Belzner}
\IEEEauthorblockA{Institute for Informatics\\
LMU Munich\\
Email: belzner@ifi.lmu.de}
\and
\IEEEauthorblockN{Thomas Gabor}
\IEEEauthorblockA{Institute for Informatics\\
LMU Munich\\
Email: belzner@ifi.lmu.de}
}


%


\maketitle

\begin{abstract}
We introduce Stacked Thompson Bandits (STB) for efficiently generating plans that are likely to satisfy a given bounded temporal logic requirement. STB uses a simulation for evaluation of plans, and takes a Bayesian approach to using the resulting information to guide its search. In particular, we show that stacking multiarmed bandits and using Thompson sampling to guide the action selection process for each bandit enables STB to generate plans that satisfy requirements with a high probability while only searching a fraction of the search space.
\end{abstract}


%
\IEEEpeerreviewmaketitle

\section{Introduction}

In many cases, system requirements can be formalized in a bounded temporal logic \cite{baier2008principles,KNP11}. For example, consider a mobile robot with planning capabilities in a area with obstacles. Here, a requirement could be that when executing a given sequential plan of ten actions, there should occur less than e.g. three collisions with obstacles. One could write this requirement as a temporal logic formula similar to the following:

$$\phi = \square_{h \leq 10}(\mathrm{collisions} \leq 2)$$

In this paper, we propose Stacked Thompson Bandits (STB) as an algorithm for constructing plans that are likely to satisfy a given requirement. STB works by simulation: It repeatedly samples a plan from its current policy, simulates the result of this plan w.r.t. a given requirement, and updates its policy according to the result in order to increase the quality of further sampled plans. As we will see, STB leverages Bayesian statistics to guide the exploration-exploitation tradeoff when searching the space of possible plans for feasible solutions.

STB is an open loop planner \cite{bubeck2010open,weinstein2013open}: It does only search the space of plans. That is, action selection is not conditioned w.r.t. intermediate states, but only w.r.t. previously selected actions. STB does only account for intermediate states when evaluating requirement satisfaction of a given plan.

The key idea of STB is to model the plan selection policy by an open loop sequence of multiarmed bandits. Each of the bandits represents an action choice for plan construction at a certain point in time. Exploration of the search space and exploitation of already gathered results are balanced with Thompson sampling \cite{thompson1933likelihood,agrawal2012analysis}. Information gathered from simulating sampled plans is used to increase the likelihood that plans constructed by STB satisfy a given requirement.

The remainder of the paper is structured as follows: Section \ref{sec:preliminaries} recaps open loop planning, multiarmed bandits and Thompson sampling. Section \ref{sec:stb} introduces Stacked Thompson Bandits. In Section \ref{sec:experiment} we discuss our empirical findings. Section \ref{sec:related} discusses related work. We conclude in Section \ref{sec:conclusion} and give pointers to potential future work.

\section{Preliminaries}
\label{sec:preliminaries}

In this Section, we recap open loop planning, multiarmed bandits and Thompson sampling.

\subsection{Open Loop Planning}
In our setting, open loop planning (\cite{bubeck2010open,weinstein2013open}) is an approach to find plans that result in satisfaction of a given goal without storing information about the intermediate states that are encountered while executing the plan. I.e. given a set of actions $A$, we are only interested in finding a plan $p \in A^*$, and we are only keeping information about the action sequences in order to guide the planning process.

This is in contrast to closed loop planning, such as e.g. Monte Carlo Tree Search \cite{chaslot2010monte}, where action selection is typically conditioned by the history of previously encountered states and executed actions.

\subsection{Multiarmed Bandits}
Multiarmed bandits (MAB) are a core framework for decision making. A bandit consists of a number of arms, each representing an agent's choice. In our setting, each arm represents an action $a \in A$. Each arms provides a particular payoff, and the agent's goal is to identify the most preferable arm. It can explore the bandit by pulling one arm at a time, and observe the corresponding payoff.

An MAB can be interpreted as a simple Markov decision process with a single state. In their basic formulation, MABs already provide a clear framework for studying the exploration-exploitation tradeoff inherent to decision making under uncertainty: Should the agent select the arm that previously showed to be most promising? Or should it go on exploring other options? 
For a recent survey of MAB and its variants, see \cite{kuleshov2014algorithms}.

\subsection{Thompson Sampling}
Thompson sampling (TS) is a Bayesian algorithm for solving an MAB. It was proposed decades ago \cite{thompson1933likelihood}, but only recently its astonishing effectiveness and generality have been identified \cite{ortega2009bayesian,chapelle2011empirical,kaufmann2012thompson}.

In the case of Bernoulli rewards (as in the case of STB), the parameter of each arm to be estimated is a probability $p \in [0;1]$. TS infers a posterior distribution over $p$ based on the observed arm payoffs and a prior assumption about the distribution of $p$. In general, the posterior is proportional to the likelihood of observed data $D$ (i.e. an arm's observed payoffs), multiplied by the prior distribution $P(\theta)$ over the parameters of interest, $\theta = p$ in our case (Equation \ref{eq:bayes}).

\begin{equation}
\label{eq:bayes}
P(\theta | D) \propto P(D | \theta) P(\theta)
\end{equation}

We can model the uncertainty about $p$ by a Beta distribution, the conjugate prior of the Bernoulli distribution. This approach ensures that the posterior is of the same form as the prior distribution, and thus enables efficient sequential updating of the distribution.
The Beta distribution is parametrized by two parameters $a, b \in \mathbb{R}^+$. In our case, $a$ and $b$ are given by the successes and failures of the arm pulls. Given $s$ successes, $f$ failures, and assuming an uninformative prior over $p$, the posterior (for $\theta = p$) is determined by Equation \ref{eq:beta}.

\begin{equation}
\label{eq:beta}
P(p | D) = \mathrm{Beta}(s + 1, f + 1)
\end{equation}

TS maintains such a distribution $P(\theta)$ for each arm. The algorithm then samples a potential value for each arm from these distributions. It then plays the arm from whose distribution the maximum value has been sampled, observes the payoff, and uses this observation to update the corresponding distribution. Repeating this process results in almost sure identification of the arm with the highest payoff. TS is schematically shown in Algorithm \ref{alg:ts}.

\begin{algorithm}
	\begin{algorithmic}[1]
		\Procedure{Thompson Sampling}{}
		\State $\forall a \in A : \hat{p}_a \sim P_a(p)$
		\State play $\arg \max_a \hat{p}_a$ and observe result
		\State update $P_a(p)$ w.r.t. result
		\EndProcedure
	\end{algorithmic}
	\caption{Thompson Sampling}
	\label{alg:ts}
\end{algorithm}

\section{Stacked Thompson Bandits}
\label{sec:stb}

We now present Stacked Thompson Bandits (STB). STB works by stacking a number of MAB, treating them as a sequential decision problem. In order to construct a plan, STB sequentially selects an action from each MAB using Thompson sampling. The resulting plan checked for requirement satisfaction using a simulation $M$. Given a set of states $S$, a set of actions $A$ and the set of bounded temporal formulae $\Phi$, $M$ is a conditional probability distribution as follows.

$$M : P(\mathrm{Bool} | S, A^*, \Phi)$$

I.e. running a simulation for a given initial state $s \in S$, a given plan $p \in A^*$ and a given requirement $\phi \in \Phi$ can be interpreted as a Bernoulli experiment. That is, $M(s, p, \phi)$ defines a Bernoulli distribution, where the result indicates whether $p$ satisfies $\phi$ when executed in $s$. STB uses the Bernoulli result of such a simulation to update each of the MABs w.r.t. Equation \ref{eq:beta}.

STB is shown in Algorithm \ref{alg:stb}. It takes the following input:
\begin{itemize}
	\item The current system state $s \in S$.
	\item A bounded temporal formula $\phi \in \Phi$.
	\item A simulation model $M$.
\end{itemize}

First, STB initializes the parameters of each stacked bandit. In particular, for each step $i$ up to the horizon of the requirement $\phi$ and each action $a \in A$, it maintains a count of successes $s_{a,i}$ (i.e. satisfactions) and failures $f_{a,i}$ (i.e. violations) (lines 2 -- 4).

Then, STB repeats the following steps until interruption (e.g. due to some budget, or because an event occurred).
\begin{itemize}
	\item An estimated satisfaction probability is sampled from the stacked bandit, i.e. from each beta distribution for each step $i$ and for each action $a$ (line 8).
	\item The actions yielding the maximum estimate for each step are combined to a plan (lines 9 and 10).
	\item The sampled plan is simulated in $M$ w.r.t. system state and given requirement. The result (satisfaction or violation) is observed (line 11).
	\item The bandit parameters of all actions in the simulated plan are updated w.r.t. the simulation result (lines 12 -- 16).
\end{itemize}

\begin{algorithm}
	\begin{algorithmic}[1]
		\Procedure {STB}{$s,\phi, M$}
		\For {$a \in A, i \in 0 ... h(\phi)$}
			\State $s_{a,i} \gets 0$
			\State $f_{a,i} \gets 0$
		\EndFor
		\While {not interrupted}
			\State $p \gets \mathrm{nil}$
			\For {$i \in 0 ... h(\phi)$}
				\State $\forall a \in A : \hat{p}_a \sim \mathrm{Beta}(s_{a,i} + 1, f_{a,i} + 1)$
				\State $a_i \gets \arg \max_{a \in A} \hat{p}_a$
				\State $p \gets p :: a_i$
			\EndFor
			\State $\mathrm{sat} \sim M(s, p, \phi)$
			\For {$a_i \in p$}
				\If {sat} \State $s_{a,i} \gets s_{a,i} + 1$
				\Else \State $f_{a,i} \gets f_{a,i} + 1$
				\EndIf
			\EndFor
		\EndWhile
		\EndProcedure
	\end{algorithmic}
	\caption{Stacked Thompson Bandits}
	\label{alg:stb}
\end{algorithm}

The algorithm terminates on some external interruption signal that is to be specified by the user. A possible decision rule on which plan to execute would be to select the action with maximum distribution mode (the value with most probability mass) for each MAB in the stack. The mode of a beta distribution with parameters $a, b$ is given by:
$$\frac{a + 1}{a + b + 2}$$

Mode plan selection for STB is therefore performed as shown in Algorithm \ref{alg:stb_mode}.

\begin{algorithm}
	\begin{algorithmic}[1]
		\For {$i \in 0 ... h$}
		\State $a_i \gets \arg \max_{a \in A} \frac{s_{a,i} + 1}{s_{a,i} + f_{a,i} + 2}$
		\State $p \gets p :: a_i$
		\EndFor
	\end{algorithmic}
	\caption{STB mode plan selection}
	\label{alg:stb_mode}
\end{algorithm}

\section{Experimental Results}
\label{sec:experiment}

We implemented STB and observed whether it is able to generate plans with increasing probability of satisfaction requirement when performing more search iterations.

\subsection{Setup}

The state $s$ is constituted by a 10 x 10 grid world, with the agent at position (0, 0). Obstacles are randomly positioned, at an obstacle to free position ratio of 0.2. Actions are movements in four directions up, down, left, right with obvious semantics.

The agent has a Bernoulli action failure probability $p_\mathrm{fail}$ uniformly sampled from $[0 ; 1]$. Action failure results in the inverse movement (e.g. failing up yields down). This constitutes domain noise in the simulation $M$ available to the agent.

The task of STB is to generate a plan of length 10 that yields less than three collisions with obstacles when executed:
$$\phi = \square_{h \leq 10}(\mathrm{collisions} \leq 2)$$

This setup yields a search space cardinality $4^{10} = 1048576$. However, note that due to probabilistic domain dynamics (i.e. the branching due to potential action failures) a plan has to be evaluated many times to obtain an adequate estimate of its satisfaction probability, yielding a very hard search problem.
We approximated the ground truth satisfaction probability of a plan by taking the maximum likelihood estimate of satisfaction probability based on 1000 simulation runs.

Our implementation of the setup and STB is available at \url{https://github.com/jazzbob/stb}.

\subsection{Results}
\begin{figure}
	\includegraphics[width=\columnwidth]{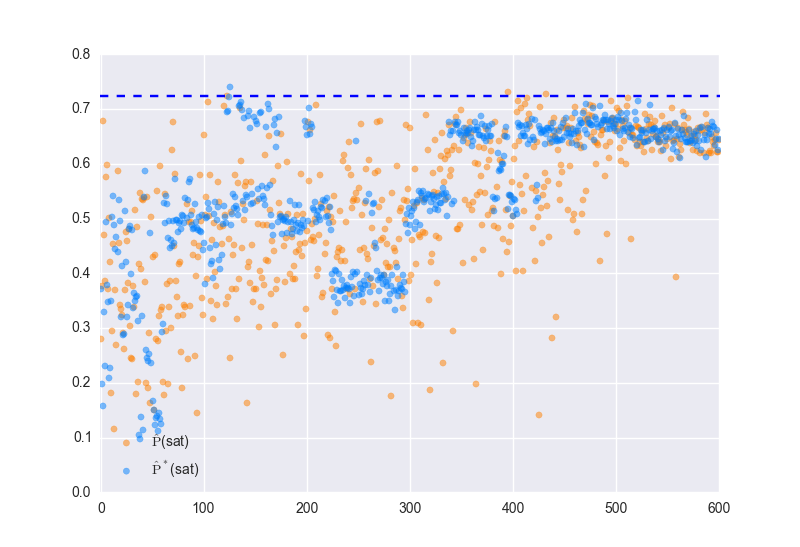}
	\caption{Exemplary STB result. The horizontal axis shows the number of search iterations. The vertical axis shows the satisfaction probability of sampled plans (orange) and of mode plans (blue) in the current iteration. Note that information about plan satisfaction probability is never explicitly available to STB. STB only uses the boolean simulation results to guide its search. The dashed line shows the optimal plan found by random search (1000 runs, 1000 evaluations each).}
	\label{fig:result}
\end{figure}

In our experimental runs, we observed that STB is able to generate plans with increasing probability of satisfying the given requirement. In particular, STB was able to find potentially close to optimal solutions searching only a fraction of the search space.
Figure \ref{fig:result} shows an exemplary run of STB.

\begin{figure}
	\includegraphics[width=\columnwidth]{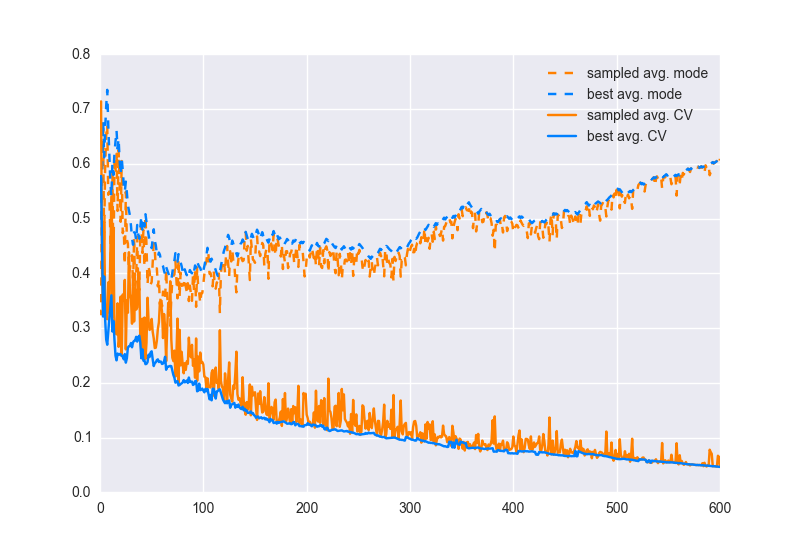}
	\caption{Exemplary STB average mode and CV for sampled and best plan respectively. The horizontal axis shows the number of search iterations. The vertical axis shows the average value of sampled (orange) and current best (blue) arm distributions' modes and CVs. As expected, average modes increase in the long run, while average CVs decrease. The strong disturbances in the beginning, in particular w.r.t. to the average modes, is due to the lack of valid information for STB to build its action value (i.e. satisfaction probability) estimates.}
	\label{fig:cv}
\end{figure}

As a sanity check, we also observed the average mode value as well as the coefficient of variation (CV) of sampled plan arm distributions and mode (i.e. best) plan arm distributions respectively. While the mode value gives a rough estimate about value of an arm (i.e. an action), the CV is defined as the ratio of standard deviation to mean $\frac{\sigma}{\mu}$ of a distribution. CV is suitable to measure the accuracy of a distribution when its mean value changes \cite{belzner2016simulation}, as is the case in STB. We expect the average mode value to increase in the long run for both sampled and best plans. We also expect the CV of sampled and best plans to decrease in the long run. We could empirically establish both expected results (cf. Figure \ref{fig:cv}).

\begin{figure}
	\includegraphics[width=\columnwidth]{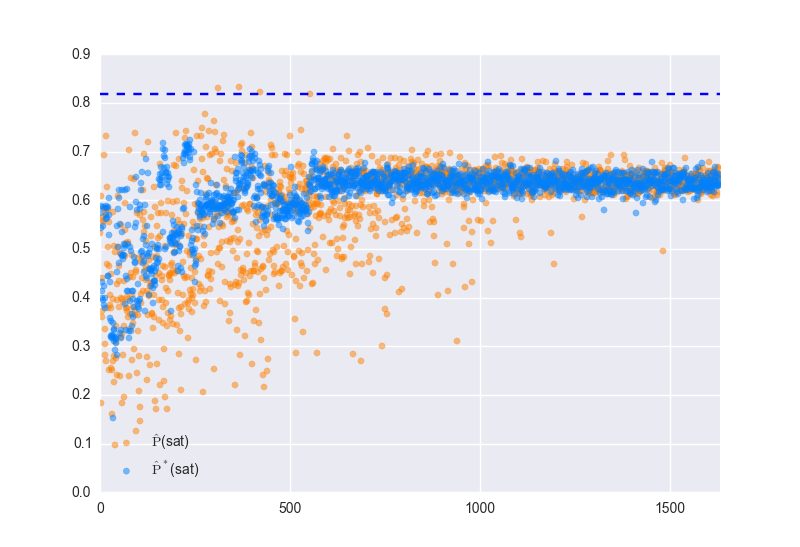}
	\caption{Exemplary run where STB is stuck in a local optimum. The horizontal axis shows the number of search iterations. The vertical axis shows the satisfaction probability of sampled plans (orange) and of mode plans (blue) in the current iteration. The dashed line shows the optimal plan found by random search (1000 runs, 1000 evaluations each).}
	\label{fig:local_opt}
\end{figure}

However, we also observed STB to get stuck in local optima (cf. Figure \ref{fig:local_opt}). This is a property of many stochastic search algorithms. One could potentially reduce the risk of getting stuck by using an ensemble of STB planners in order to reduce the probability of such premature convergence.

\section{Related Work}
\label{sec:related}

Our work on STB is strongly influenced by existing open loop planners. In particular, Cross Entropy Open Loop Planning is an approach for planning in large-scale continuous MDPs \cite{weinstein2013open}. It is however not applicable to discrete domains as STB. Recently, cross entropy planning has been used for searching sequences that satisfy a given temporal logic formula \cite{livingston2015cross} in a continuous motion planning setting.

STB is subtly related to statistical model checking (SMC) \cite{younes2006numerical,legay2010statistical}, and Bayesian statistical model checking in particular \cite{langmead2009generalized,jha2009bayesian,zuliani2010bayesian}. Here, the setting is to guarantee a minimal required satisfaction probability for a given, fixed sequence of actions. SMC approaches are able to provide such a result, potentially with a quantifiable confidence. STB is not able to provide a quantification of satisfaction probability, as is samples every plan only once. Also, the distributions maintained in the stacked bandits to not provide an accurate absolute quantification due to the sampling nature of search and the corresponding concept drift. On the other hand, STB is able to generate plans with high satisfaction probability, whereas SMC only can determine this probability. In practice, a combination of both approaches is possible and seems useful.

\section{Conclusion}
\label{sec:conclusion}

We have presented Stacked Thompson Bandits (STB), an open loop planning algorithm for generating action sequences that satisfy a given bounded temporal logic requirement with high probability. STB works by maintaining a stack of multiarmed bandits via Thompson sampling. We have preliminarily and empirically evaluated the effectiveness of STB on a toy example.

There are various directions for future work. As mentioned, it would be interesting to see if an ensemble approach with STB could reduce the probability of getting stuck in local minima. Another interesting venue would be to combine temporal action abstraction with STB. See \cite{belzner2016time, belzner2016simulation} for previous work of one of the authors on temporal abstraction in open loop planning. Also, guiding the sampling process in a QoS-aware manner based on required confidences could prove worthwhile. See e.g. \cite{belzner2016qos} for previous work of the authors on QoS-aware sampling in multiarmed bandits. It would also be interesting to explore STB under model uncertainty, e.g. when the simulation model is learned from incomplete information.




\bibliographystyle{IEEEtran}
\bibliography{IEEEabrv,references}
%
%
%

\end{document}